\documentclass[aps,twocolumn,showpacs,preprintnumbers,floats]{revtex4}\def\@cite#1#2{\textsuperscript{[{#1\if@tempswa , #2\fi}]}}
\usepackage{mathrsfs}
\usepackage{longtable,lscape}
\usepackage{txfonts}
\usepackage{amssymb}
\usepackage{indentfirst}
\usepackage{graphicx,booktabs}

\usepackage{multirow}
\usepackage{color}
\usepackage{epsfig}
\newcommand{\vsig}{\mbox{\boldmath$\sigma$\unboldmath}}

\begin{document}

\title{Strong and radiative decays of the doubly charmed baryons}

\author{Li-Ye Xiao$^{1,2}$~\footnote {E-mail: lyxiao@pku.edu.cn}, Kai-Lei Wang$^{3}$, Qi-Fang L$\ddot{\text{u}}$$^{3,4}$, Xian-Hui Zhong$^{3,4}$~\footnote {E-mail: zhongxh@hunnu.edu.cn}, and Shi-Lin Zhu$^{1,2,5}$~\footnote {E-mail: zhusl@pku.edu.cn}}

\affiliation{ 1) School of Physics and State Key Laboratory of
Nuclear Physics and Technology, Peking University, Beijing 100871,
China }
\affiliation{ 2)  Center of High
Energy Physics, Peking University, Beijing 100871, China}
\affiliation{ 3) Department of Physics, Hunan Normal
University, and Key Laboratory of Low-Dimensional Quantum Structures
and Quantum Control of Ministry of Education, Changsha 410081, China
} \affiliation{ 4) Synergetic Innovation Center for Quantum Effects
and Applications (SICQEA), Hunan Normal University, Changsha 410081,
China} \affiliation{ 5)  Collaborative Innovation Center of Quantum
Matter, Beijing 100871, China}

%\date{\today}

\begin{abstract}

We have systematically studied the strong and radiative decays of
the low-lying $1P$-wave doubly charmed baryons. Some interesting
observations are: (i) The states $\Xi_{cc}^*$ and $\Omega_{cc}^*$
with $J^P=3/2^+$ have a fairly large decay rate into the
$\Xi_{cc}\gamma$ and $\Omega_{cc}\gamma$ channels with a width $\sim
15$ and $\sim7$ keV, respectively. (ii) The lowest lying excited
doubly charmed baryons are dominated by the $1P$ $\rho$ mode excitations, which
should be quite narrow states. They decay into the ground
state with $J^P=1/2^+$ through the radiative transitions with a significant ratio. (iii) The
total decay widths of the first orbital excitations of $\lambda$
mode ($1P_{\lambda}$ states with $J^P=1/2^-$, $3/2^-$, $5/2^-$) are
about $\Gamma\sim100$ MeV, and the ratio between the radiative and
hadronic decay widths is about $\mathcal{O}(10^{-3})$.

\end{abstract}


\maketitle

\section{Introduction}

In the past three decades, many singly heavy baryons were observed
experimentally~\cite{Olive:2016xmw}. However, the experimental
progress on the doubly heavy baryons remains very challenging. The SELEX
Collaboration reported some evidence of two signals
$\Xi^+_{cc}(3519)$~\cite{Mattson:2002vu} and $\Xi^{++}_{cc}(3770)$~\cite{Moinester:2002uw}, which were not confirmed by other
collaborations unfortunately. Very recently, the doubly heavy baryon
$\Xi^{++}_{cc}(3621)$ was discovered in the
$\Lambda^+_cK^-\pi^+\pi^+$ mass spectrum by the LHCb
collaboration~\cite{Aaij:2017ueg}. Its mass was measured to be
3621.40$\pm$0.72(stat)$\pm$0.27(syst)$\pm0.14$ MeV. The newly
observed $\Xi_{cc}(3621)^{++}$ has attracted a great deal of
attention from the hadron physics community
\cite{n1,n2,Wang:2017mqp,n4,n5,n6,n7,n8,Li:2017,Yu:2017,Hu:2005gf}.

The doubly heavy baryons provide a new platform to study the heavy
quark symmetry and chiral dynamics simultaneously. There exist many
theoretical calculations of the mass spectra of the doubly charmed
baryons with various models in the literature
~\cite{Roncaglia:1995az,Gershtein:2000nx,Itoh:2000um,Ebert:2002ig,Roberts:2007ni,Yoshida:2015tia,Shah:2016vmd,Gershtein:1998sx,Zhang:2008rt,Wang:2010hs,Brown:2014ena}.
The semi-leptonic decays of the doubly charmed baryons were also
studied extensively
~\cite{Faessler:2001mr,Albertus:2006ya,Roberts:2008wq,Albertus:2009ww,Faessler:2009xn,Onishchenko:2000wf,Kiselev:2001fw,White:1991hz,SanchisLozano:1994vh,Hernandez:2007qv,
Guo:1998yj,Ebert:2004ck,Wang:2017mqp}. Furthermore, there are a few discussions of
the radiative transitions of the doubly charmed baryons in the
literature ~\cite{Hackman:1977am,Branz:2010pq,Bernotas:2013eia,Li:2017}. In
this work, we shall perform a systematical investigation of both
strong and radiative decays of the low-lying $1P$-wave
doubly charmed baryons. Their quark model classification, their
allowed decay channels, and their predicted masses from Ref.~\cite{Ebert:2002ig}
are summarized in Table~\ref{mass}.

To deal with the pionic or kaonic decays of the doubly charmed baryons, we
apply the chiral quark model~\cite{Manohar:1983md}, which was quite
successful in the description of the hadronic decays of the
heavy-light mesons and baryons
~\cite{Zhong:2007gp,Zhong:2009sk,Zhong:2008kd,Zhong:2010vq,Xiao:2014ura,
Xiao:2013xi,Liu:2012sj,Nagahiro:2016nsx,Wang:2017hej} and light
pseudoscalar meson
productions~\cite{Li:1995si,Li:1998ni,Li:1995vi,Zhao:2002id,Li:1994cy,Li:1997gd,Saghai:2001yd,Zhao:2000iz,He:2008ty,He:2008uf,Zhong:2007fx,Zhong:2008km,Zhong:2009zz,
Zhong:2011ti,Zhong:2011ht,Zhong:2013oqa,Xiao:2013hca,Xiao:2015gra,Xiao:2016dlf,Wang:2017cfp}.
The radiative decays of the doubly charmed baryons are analyzed
within the constituent quark model. The same formalism was
successfully applied to study the radiative decays of the $c\bar{c}$
and $b\bar{b}$ systems ~\cite{Deng:2016stx,Deng:2016ktl} and the
newly observed $\Omega_c$ states~\cite{Wang:2017hej}.

The paper is structured as follows. In Sec. II we review the quark
model description of the strong and radiative decays of the $ccq$
system. We present the numerical results and some discussions in
Sec. III and summarize our results in Sec. IV.

\begin{table*}
\caption{\label{mass} Masses and possible two body decay channels of the $1P$ doubly
charmed baryons ( denoted by $|N^{2S+1}L_{\sigma}J^P\rangle$), where
$|N^{2S+1}L_{\sigma}J^P\rangle$=$\sum_{L_z+S_z=J_z}\langle
LL_z,SS_z|JJ_z\rangle^N\Psi^{\sigma}_{LL_z}\chi_{S_z}\phi$
~\cite{Zhong:2007gp}. The masses (MeV) are taken from the relativistic quark model~\cite{Ebert:2002ig}. }
\begin{tabular}{ccccccccccc}\hline\hline
State  & & \multicolumn{3}{c}{$\Xi_{cc}$} &\multicolumn{3}{c}{$\Omega_{cc}$} \\ \cline{3-5}\cline{7-9}
$N^{2S+1}L_{\sigma}J^P$  ~~& Wave function ~~&Mass~\cite{Ebert:2002ig}~~~&Decay channel &Observed state & &Mass~\cite{Ebert:2002ig}~~~&Decay channel &Observed state  \\
 $|0^2S\frac{1}{2}^+\rangle$  &$^{0}\Psi^S_{00}\chi^\lambda_{S_z}\phi$   &3620~~~  &   & $\Xi(3621)$\cite{Aaij:2017ueg}?         &&3778 ~~~&  &?\\
 $|0^4S\frac{3}{2}^+\rangle$  &$^{0}\Psi^S_{00}\chi^s_{S_z}\phi$         &3727~~~  &$\Xi_{cc}\gamma$ & ?                         &&3872~~~
 &$\Omega_{cc}\gamma$   &?           \\
 $|1^2P_{\rho}\frac{1}{2}^-\rangle$&$^{1}\Psi^{\rho}_{1L_z} \chi^{\rho}_{S_z}\phi $ & 3838 ~~~&$\Xi_{cc}\gamma$, $\Xi^*_{cc}\gamma$ & ?   &&4002~~~& $\Omega_{cc}\gamma$,$\Omega_{cc}^*\gamma$ &?           \\
 $|1^2P_{\rho}\frac{3}{2}^-\rangle$&         & 3959 ~~~&$\Xi_{cc}\gamma$, $\Xi^*_{cc}\gamma$ & ?        &&4102~~~&$\Omega_{cc}\gamma$,$\Omega_{cc}^*\gamma$  &?   \\
 $|1^2P_{\lambda}\frac{1}{2}^-\rangle$ &$^{1}\Psi^{\lambda}_{1L_z} \chi^{\lambda}_{S_z}\phi$ & 4136 ~~~&$\Xi_{cc}\pi$, $\Xi^*_{cc}\pi$,$\Xi_{cc}\gamma$, $\Xi^*_{cc}\gamma$& ?   &&4271~~~&$\Xi_{cc}K$, $\Xi_{cc}^*K$, $\Omega_{cc}\gamma$, $\Omega_{cc}^*\gamma$ &?              \\
 $|1^2P_{\lambda}\frac{3}{2}^-\rangle$  &    & 4196~~~&$\Xi_{cc}\pi$, $\Xi^*_{cc}\pi$,$\Xi_{cc}\gamma$, $\Xi^*_{cc}\gamma$ & ?   &&4325~~~&$\Xi_{cc}K$, $\Xi_{cc}^*K$, $\Omega_{cc}\gamma$, $\Omega_{cc}^*\gamma$ & ?        \\
 $|1^4P_{\lambda}\frac{1}{2}^-\rangle$  &$^{1}\Psi^{\lambda}_{1L_z} \chi^{s}_{S_z}\phi$  & 4053 ~~~&$\Xi_{cc}\pi$, $\Xi^*_{cc}\pi$,$\Xi_{cc}\gamma$, $\Xi^*_{cc}\gamma$ & ?   &&4208~~~& $\Xi_{cc}K$, $\Xi_{cc}^*K$, $\Omega_{cc}\gamma$, $\Omega_{cc}^*\gamma$  & ?            \\
 $|1^4P_{\lambda}\frac{3}{2}^-\rangle$ &                 & 4101 ~~~&$\Xi_{cc}\pi$, $\Xi^*_{cc}\pi$,$\Xi_{cc}\gamma$, $\Xi^*_{cc}\gamma$& ?   &&4252 ~~~ & $\Xi_{cc}K$, $\Xi_{cc}^*K$, $\Omega_{cc}\gamma$, $\Omega_{cc}^*\gamma$ & ?          \\
 $|1^4P_{\lambda}\frac{5}{2}^-\rangle$&      & 4155~~~ &$\Xi_{cc}\pi$, $\Xi^*_{cc}\pi$,$\Xi_{cc}\gamma$, $\Xi^*_{cc}\gamma$ & ?   &&4303 ~~~& $\Xi_{cc}K$, $\Xi_{cc}^*K$, $\Omega_{cc}\gamma$, $\Omega_{cc}^*\gamma$ & ?           \\
\hline\hline
\end{tabular}
\end{table*}
%\end{center}
%\end{widetext}

\section{Chiral quark model}\label{model}

In the chiral quark model, the effective low energy
quark-pseudoscalar-meson coupling in the SU(3) flavor basis at tree
level is described by ~\cite{Manohar:1983md}
 \begin{eqnarray}\label{STcoup}
H_m=\sum_j
\frac{1}{f_m}\bar{\psi}_j\gamma^{j}_{\mu}\gamma^{j}_{5}\psi_j\partial^{\mu}\phi_m,
\end{eqnarray}
where $\psi_j$ stands for the $j$-th quark field in a baryon. $f_m$
is the pseudoscalar meson decay constant and $\phi_m$ is the
pseudoscalar meson octet
 \begin{eqnarray}
\phi_m=\pmatrix{
 \frac{1}{\sqrt{2}}\pi^0+\frac{1}{\sqrt{6}}\eta & \pi^+ & K^+ \cr
 \pi^- & -\frac{1}{\sqrt{2}}\pi^0+\frac{1}{\sqrt{6}}\eta & K^0 \cr
 K^- & \bar{K}^0 & -\sqrt{\frac{2}{3}}\eta}.
\end{eqnarray}

For the radiative decay under this model framework, we adopt the
quark-photon electromagnetic interaction ~\cite{Brodsky:1968ea}:
\begin{eqnarray}\label{EMcoup}
H_e=-\sum_je_j\bar{\psi}_j\gamma^{j}_{\mu}A^{\mu}(\mathbf{k},\mathbf{r_j})\psi_j,
\end{eqnarray}
where $\mathbf{k}$ stands for the three-momentum of the photon with
the field $A^{\mu}$. $\mathbf{r}_j$ and $e_j$ represent the
coordinate and charge of the $j$-th constituent quark, respectively.
This model was successfully applied to discuss the radiative decay
of doubly heavy mesons~\cite{Deng:2016stx,Deng:2016ktl}.

\begin{figure}[]
\centering \epsfxsize=3.0 cm \epsfbox{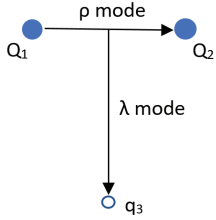} \caption{(Color
online) The $\rho-$ and $\lambda-$ mode excitations of the $ccq$
system where $\rho$ and $\lambda$ are the Jacobi coordinates defined
as $\rho=\frac{1}{\sqrt{2}}(\mathbf{r}_1-\mathbf{r}_2)$ and
$\lambda=\frac{1}{\sqrt{6}}(\mathbf{r}_1+\mathbf{r}_2-2\mathbf{r}_3)$,
respectively. $Q_1$ and $Q_2$ stand for the charm quark, and $q_3$
stands for the light ($u, d, s)$ quark. }\label{fig-str}
\end{figure}

To match the nonrelativistic harmonic oscillator spatial wave
function $^N\Psi_{LL_z}$ in this work, we adopt a nonrelativistic
form of the quark-pseudoscalar and quark-photon EM couplings. The
nonrelativistic form of Eq.~(\ref{STcoup}) reads
~\cite{Zhao:2002id,Li:1994cy,Li:1997gd}
  \begin{eqnarray}\label{non-relativistic-expansST}
H^{nr}_{m}&=&\sum_j\Big\{\frac{\omega_m}{E_f+M_f}\vsig_j\cdot
\textbf{P}_f+ \frac{\omega_m}{E_i+M_i}\vsig_j \cdot
\textbf{P}_i  \\
&&-\vsig_j \cdot \textbf{q} +\frac{\omega_m}{2\mu_q}\vsig_j\cdot
\textbf{p}'_j\Big\}I_j \phi_m,
\end{eqnarray}
where the $\vsig_j$ and $\mu_q$ stands for the Pauli spin vector and
the reduced mass of the $j$-th quark in the initial and final
baryons, respectively. $\varphi_m=e^{-i\textbf{q}\cdot
\textbf{r}_j}$ for emitting a meson, and
$\varphi_m=e^{i\textbf{q}\cdot \textbf{r}_j}$ for absorbing  a
meson. $\textbf{p}'_j=\textbf{p}_j-(m_j/M) \textbf{P}_{c.m.}$ is the
internal momentum of the $j$-th quark in the baryon rest frame.
$\omega_m$ and $\textbf{q}$ are the energy and three-vector momentum
of the meson, respectively.  $I_j$ is the isospin operator
associated with the pseudoscalar meson
\begin{eqnarray}
I_j=\cases{ a^{\dagger}_j(u)a_j(s) & for $K^-$,\cr
a^{\dagger}_j(d)a_j(s) & for $K^0$, \cr
 a^{\dagger}_j(u)a_j(d)  &
for $\pi^-$,\cr
 a^{\dagger}_j(d)a_j(u)  &
for $\pi^+$,\cr
\frac{1}{\sqrt{2}}[a^{\dagger}_j(u)a_j(u)-a^{\dagger}_j(d)a_j(d)] &
for $\pi^0$.}
\end{eqnarray}

The nonrelativistic form of Eq.~(\ref{EMcoup}) reads
~\cite{Zhao:2002id,Li:1994cy,Li:1997gd,Deng:2016stx,Deng:2016ktl,Brodsky:1968ea}
\begin{eqnarray}\label{non-relativistic-expansEM}
H^{nr}_{e}&=&\sum_j\left[e_j\mathbf{r}_j\cdot\mathbf{\epsilon}-\frac{e_j}{2m_j}\sigma_j\cdot(\mathbf{\epsilon}\times
\mathbf{\hat{k}})\right]e^{-i\mathbf{k}\cdot \mathbf{r}_j},
\end{eqnarray}
where the $\mathbf{\epsilon}$ is the polarization vector of the
photon.

\begin{figure}[b]
\centering \epsfxsize=7cm \epsfbox{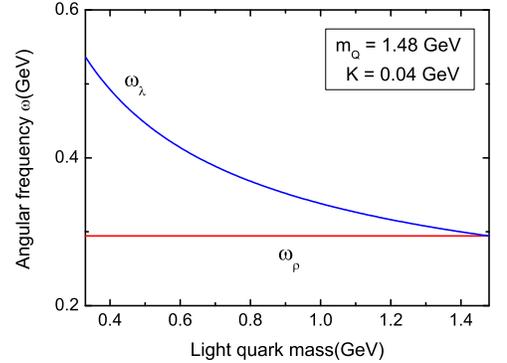} \caption{(Color
online) Light-quark mass dependence of the excitation energy of the
$\lambda$ mode (blue solid line) and the $\rho$ mode (red solid
line) in Eq.~(\ref{os}).}\label{fig-rad}
\end{figure}

For the emission of a light pseudoscalar meson, the partial decay
width is
\begin{equation}
\Gamma_m=(\frac{\delta}{f_m})^2\frac{(E_f +M_f)|q|}{4\pi
M_i}\frac{1}{2J_i+1}\sum_{J_{iz}J_{fz}}^{}|M_{J_{iz},J_{fz}}|^2,
\end{equation}
where $M_{J_{iz},J_{fz}}$ is the transition amplitude, $J_{iz}$ and
$J_{fz}$ stand for the third components of the total angular momenta
of the initial and final baryons, respectively. Accounting for the
strength of the quark-meson coupling, $\delta$ is a global parameter
which has been determined in previous study of the strong decays of
the charmed baryons and heavy-light
mesons~\cite{Zhong:2007gp,Zhong:2008kd}. Here, we fix its value the
same as that in Refs.~\cite{Zhong:2007gp,Zhong:2008kd}, i.e.,
$\delta=0.557$.

Meanwhile, the partial radiative decay widths are
~\cite{Deng:2016stx,Deng:2016ktl}
\begin{equation}
\Gamma_{\gamma}=\frac{|\mathbf{k}|^2}{\pi}\frac{2}{2J_i+1}\frac{M_f}{M_i}\sum_{J_{iz}J_{fz}}|A_{J_{iz},J_{fz}}|^2,
\end{equation}
where $A_{J_{iz},J_{fz}}$ stands for the EM transition amplitude.

In the calculation, the standard quark model parameters are adopted.
Namely, we set $m_u=m_d=330$ MeV, $m_s=450$ MeV, and $m_c=1480$ MeV
for the constituent quark masses. The decay constants for $\pi$ and
$K$ mesons are taken as $f_{\pi}=132$ MeV, $f_{K}=160$ MeV,
respectively. The masses of the resonances are then from the predictions
with the relativistic quark model~\cite{Ebert:2002ig}. The mass of
the ground-state $|\Xi_{cc}$$ ^2S\frac{1}{2}^+\rangle$ is adopted the
experimental measurement, $M=3621$ MeV. The harmonic oscillator
parameter $\alpha_{\rho}$ in the spatial wave function of the
$\rho$-mode excitation between the two charm quarks is taken as
$\alpha_{\rho}=0.66$ GeV as in the charmonium system, which is
significantly larger than that of the $\rho$-mode excitation between
the two strange quarks ($\alpha_{\rho}=0.44$ GeV) adopted in~\cite{Wang:2017hej}. Another harmonic oscillator
parameter $\alpha_{\lambda}$ is estimated with the relation:
\begin{equation}\label{aa}
\alpha_\lambda=\left({3m_q\over
2m_Q+m_q}\right)^{1/4}\alpha_{\rho}.
\end{equation}

In the simplified case of the harmonic oscillator potential, the
$\rho$ and $\lambda$ degrees of freedom decouple,
\begin{equation}\label{Ha}
H=\frac{1}{2m_{\rho}}\mathbf{p}_{\rho}^2+\frac{1}{2m_{\lambda}}\mathbf{p}_{\lambda}^2+\frac{3}{2}K(\rho^2+\lambda^2),
\end{equation}
where
\begin{equation}
m_{\rho}=m_Q, ~~~~~~~~~m_{\lambda}=\frac{3m_Qm_q}{2m_Q+m_q}
\end{equation}
and the oscillator frequencies $\omega_{\rho}$ and
$\omega_{\lambda}$ are defined as
\begin{equation}\label{os}
\omega_{\rho}=(3K/m_{\rho})^{1/2},
~~~~~~~~~\omega_{\lambda}=(3K/m_{\lambda})^{1/2}.
\end{equation}
The ratio of the $1P$ $\rho$ and $\lambda$ excitation energies reads
\begin{equation}
\frac{\omega_{\lambda}}{\omega_{\rho}}=\sqrt{\frac{1}{3}+\frac{2m_Q}{3m_q}}>1.
\end{equation}
Since the bottom and charm quark masses are much larger than the
light quark mass ($m_Q>m_q$), the excitation energy of the
$\lambda$-mode is larger than that of the $\rho$-mode,
$\omega_{\lambda}>\omega_{\rho}$ (see Fig.~\ref{fig-rad}). Thus, the
$\rho$-excitation modes are lighter than the $\lambda$-excitation
modes for the $1P$ doubly charmed baryons. The realistic potential
is much more complicated than the simple harmonic oscillator
potential. However, the general feature of the level ordering of the
$1P$ doubly charmed baryons should be similar.

\begin{table}[]
\caption{Radiative decay widths (keV) of the ground states with $J^P=3/2^+$
compared with the three-quark model~\cite{Branz:2010pq} and chiral perturbation theory~\cite{Li:2017}.  } \label{Srad}
\begin{tabular}{cccccccc}\hline\hline
Process \ \ \ \ \ \  \ \ \ & Our result \ \ \ \ \ \  \ \ \ & \cite{Branz:2010pq}\ \ \ \ \ \  \ \ \ & \cite{Li:2017} \\
$\Xi_{cc}^{*++}\rightarrow \Xi_{cc}^{++}$ \ \ \ \ \ \  \ \ \ & 16.7 \ \ \ \ \ \  \ \ \ & 23.5 \ \ \ \ \ \  \ \ \ & 22.0 \\
$\Xi_{cc}^{*+}\rightarrow \Xi_{cc}^{+}$ \ \ \ \ \ \  \ \ \ & 14.6 \ \ \ \ \ \  \ \ \ & 28.8 \ \ \ \ \ \  \ \ \ & 9.57 \\
$\Omega_{cc}^{*}\rightarrow \Omega_{cc}$ \ \ \ \ \ \  \ \ \ & 6.93 \ \ \ \ \ \  \ \ \ & 2.11 \ \ \ \ \ \  \ \ \ & 9.45 \\
$\Xi_{bb}^{*0}\rightarrow \Xi_{bb}^{0}$ \ \ \ \ \ \  \ \ \ & 1.19 \ \ \ \ \ \  \ \ \ & 0.31 \ \ \ \ \ \  \ \ \ &$\cdot\cdot\cdot$  \\
$\Xi_{bb}^{*-}\rightarrow \Xi_{bb}^{-}$ \ \ \ \ \ \  \ \ \ & 0.24 \ \ \ \ \ \  \ \ \ & 0.06 \ \ \ \ \ \  \ \ \ &$\cdot\cdot\cdot$  \\
$\Omega_{bb}^{*}\rightarrow \Omega_{bb}$ \ \ \ \ \ \  \ \ \ & 0.08 \ \ \ \ \ \  \ \ \ & 0.02 \ \ \ \ \ \  \ \ \ &$\cdot\cdot\cdot$  \\
\hline\hline
\end{tabular}
\end{table}

\begin{figure*}[]
\centering \epsfxsize=15.0 cm \epsfbox{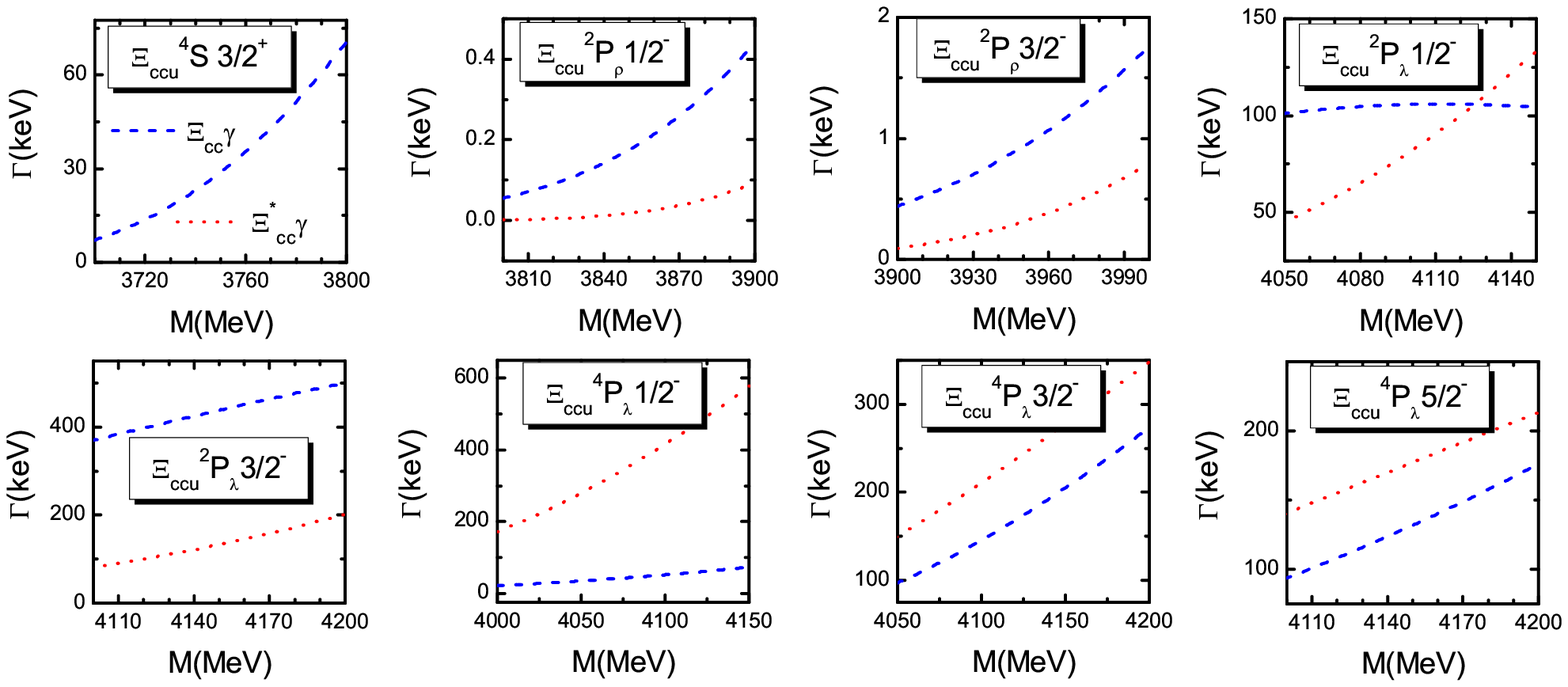}
\caption{(Color online) The radiative decay widths of the low-lying
$S$- and $P$-wave $\Xi_{ccu}$ states as a function of the mass. In
the figure, $\Xi_{cc}$ and $\Xi^*_{cc}$ stand for the
$|\Xi_{ccu}$$^2S\frac{1}{2}^+\rangle$ and
$|\Xi_{ccu}$$^4S\frac{3}{2}^+\rangle$ states. Thir masses are 3621
MeV and 3727 MeV, respectively. }\label{fig-Xpr1}
\end{figure*}

\begin{figure*}[]
\centering \epsfxsize=15.0 cm \epsfbox{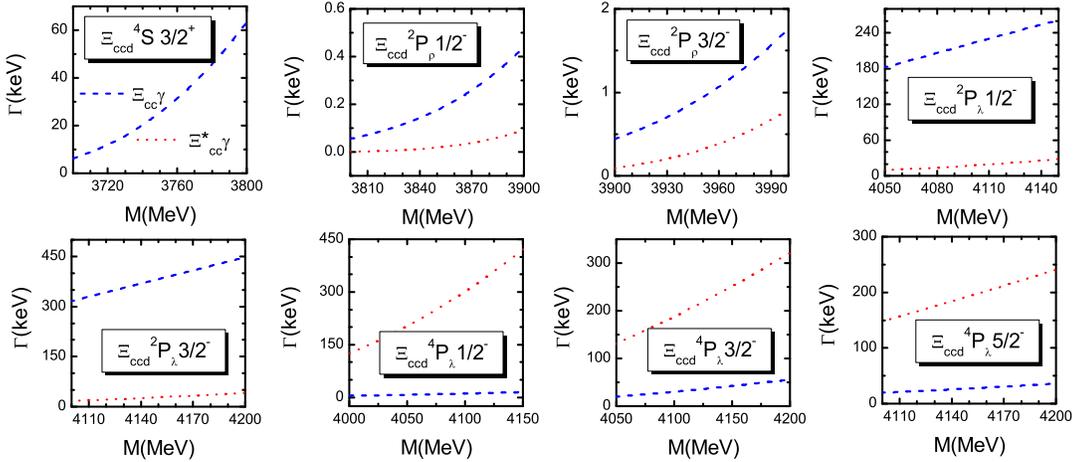}
\caption{(Color online)  The radiative decay widths of the low-lying
$S$- and $P$-wave $\Xi_{ccd}$ states as a function of the mass.
}\label{fig-Xpr2}
\end{figure*}
\begin{figure*}[]
\centering \epsfxsize=15.0 cm \epsfbox{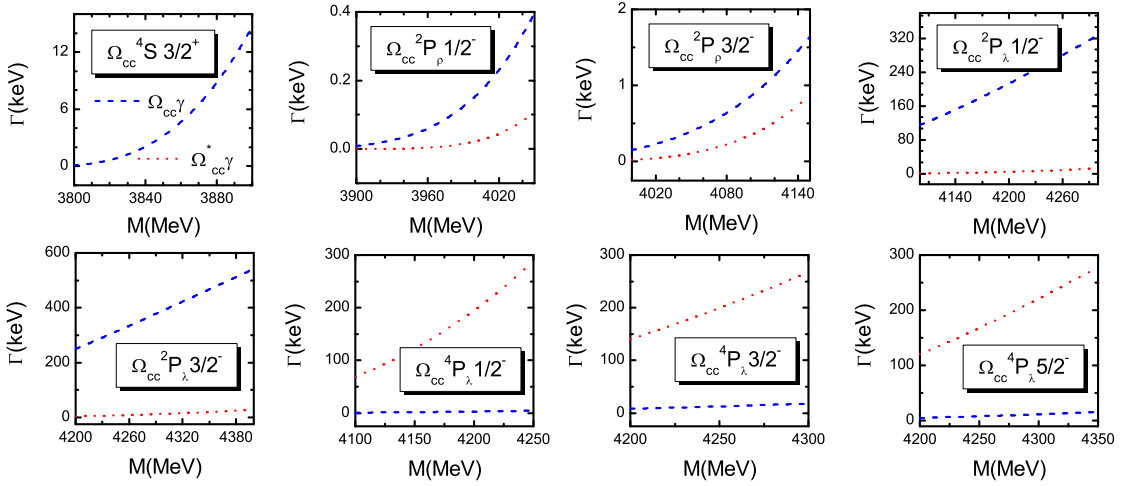}
\caption{(Color online) The radiative decay widths of the low-lying $S$- and
$P$-wave $\Omega_{cc}$ states as a function of the mass. In the
figure, $\Omega_{cc}$ and $\Omega^*_{cc}$ stand for the
$|\Omega_{cc}$$^2S\frac{1}{2}^+\rangle$ and
$|\Omega_{cc}$$^4S\frac{3}{2}^+\rangle$ states.  Their masses are
3778 MeV  and 3872 MeV, respectively. }\label{fig-Opr}
\end{figure*}

%\begin{widetext}
%\begin{center}
\begin{table*}[!htbp]
\caption{The partial widths of strong and radiative decays for the $1P_{\lambda}$ states.
$\Gamma_{\text{total}}$ stands for the total decay width.  } \label{PLV}
\begin{tabular}{cccccccc}\hline\hline
 \ \ \ \ \ \  \ \ \ &\ \ \ \ \ \ \ &     \ \ \ &   \ \ \ & \text{Types of}\ \ \ \ \ \ \ \ \ \ \ \ & \ \ \ \  \ \ \ &   \ \ \ \    \ \ \ &       \\
State \ \ \ \ \ \    &  Mass(MeV)  \ \ \ \ \  \  & $\Gamma[\Xi_{cc}\pi]$(MeV) \ \ \ \ \  \  & $\Gamma[\Xi^*_{cc}\pi]$(MeV)  \ \ \ \ \  \  & \text{light quark}   \ \ \ \ \   \  & $\Gamma[\Xi_{cc}\gamma]$(keV)  \ \ \ \ \   \  & $\Gamma[\Xi^*_{cc}\gamma]$(keV)   \ \ \ \ \   \  & $\Gamma_{\text{total}}$(MeV)      \\
$|\Xi_{cc}$$^2P_{\lambda}\frac{1}{2}^-\rangle$ \ \ \ \ \ \  \ \ \ & 4136 \ \ \ \ \ \     \ \ \ &15.6  \ \ \ \ \ \     \ \ \  &33.9 \ \ \ \ \ \     \ \ \  &$u$  \ \ \ \ \ \     \ \ \  &105 \ \ \ \ \ \     \ \ \  &117   \ \ \ \ \ \     \ \ \ &49.7  \\
                                                 \ \ \ \ \ \     \ \ \  &        \ \ \ \ \ \     \ \ \  &   \ \ \ \ \ \     \ \ \  &   \ \ \ \ \ \     \ \ \  &$d$  \ \ \ \ \ \     \ \ \  &250  \ \ \ \ \ \     \ \ \ &24.6  \ \ \ \ \ \     \ \ \  &     \\
$|\Xi_{cc}$$^2P_{\lambda}\frac{3}{2}^-\rangle$  \ \ \ \ \ \     \ \ \ & 4196 \ \ \ \ \ \     \ \ \ &21.6  \ \ \ \ \ \     \ \ \ &101  \ \ \ \ \ \     \ \ \  &$u$  \ \ \ \ \ \     \ \ \ &495  \ \ \ \ \ \     \ \ \  &196   \ \ \ \ \ \     \ \ \  &123  \\
                                                 \ \ \ \ \ \     \ \ \  &        \ \ \ \ \ \     \ \ \  &   \ \ \ \ \ \     \ \ \  &  \ \ \ \ \ \     \ \ \  &$d$  \ \ \ \ \ \     \ \ \  &442   \ \ \ \ \ \     \ \ \ &40.7  \ \ \ \ \ \     \ \ \  &       \\
$|\Xi_{cc}$$^4P_{\lambda}\frac{1}{2}^-\rangle$  \ \ \ \ \ \     \ \ \ & 4053  \ \ \ \ \ \     \ \ \  &133  \ \ \ \ \ \     \ \ \ &1.22  \ \ \ \ \ \     \ \ \  &$u$   \ \ \ \ \ \     \ \ \ &35.7  \ \ \ \ \ \     \ \ \  &287   \ \ \ \ \ \     \ \ \ &134  \\
                                               \ \ \ \ \ \     \ \ \  &    \ \ \ \ \ \     \ \ \  &   \ \ \ \ \ \     \ \ \  &   \ \ \ \ \ \     \ \ \  &$d$  \ \ \ \ \ \     \ \ \  &7.47 \ \ \ \ \ \     \ \ \  &208  \ \ \ \ \ \     \ \ \ &      \\
$|\Xi_{cc}$$^4P_{\lambda}\frac{3}{2}^-\rangle$  \ \ \ \ \ \     \ \ \  & 4101 \ \ \ \ \ \     \ \ \ &7.63    \ \ \ \ \ \     \ \ \  &84.6   \ \ \ \ \ \     \ \ \  &$u$  \ \ \ \ \ \     \ \ \  &147  \ \ \ \ \ \     \ \ \  &212  \ \ \ \ \ \     \ \ \  & 92.6  \\
                                                \ \ \ \ \ \     \ \ \  &    \ \ \ \ \ \     \ \ \  &     \ \ \ \ \ \     \ \ \  &     \ \ \ \ \ \     \ \ \ &$d$  \ \ \ \ \ \     \ \ \  &30.5  \ \ \ \ \ \     \ \ \ &189  \ \ \ \ \ \     \ \ \ &      \\
$|\Xi_{cc}$$^4P_{\lambda}\frac{5}{2}^-\rangle$  \ \ \ \ \ \     \ \ \  & 4155 \ \ \ \ \ \     \ \ \ &75.3  \ \ \ \ \ \     \ \ \ &22.8   \ \ \ \ \ \     \ \ \ &$u$   \ \ \ \ \ \     \ \ \  &136   \ \ \ \ \ \     \ \ \  &181   \ \ \ \ \ \     \ \ \  &98.4  \\
                                                \ \ \ \ \ \     \ \ \  &  \ \ \ \ \ \     \ \ \  &    \ \ \ \ \ \     \ \ \  &     \ \ \ \ \ \     \ \ \  &$d$   \ \ \ \ \ \     \ \ \  &28.0 \ \ \ \ \ \     \ \ \ &198    \ \ \ \ \ \     \ \ \  &      \\
\hline\hline
State \ \ \ \ \ \   &  Mass(MeV)  \ \ \  \ \ \ & $\Gamma[\Xi_{cc}K]$(MeV)  \ \ \  \ \ \ & $\Gamma[\Xi^*_{cc}K]$(MeV)  \ \ \  \ \ \ &   \ \ \ \ \ \   \ \ \ & $\Gamma[\Omega_{cc}\gamma]$(keV)  \ \ \   \ \ \ & $\Gamma[\Omega^*_{cc}\gamma]$(keV)  \ \ \   \ \ \ & $\Gamma_{\text{total}}$(MeV)      \\
$|\Omega_{cc}$$^2P_{\lambda}\frac{1}{2}^-\rangle$  \ \ \ \ \ \     \ \ \  & 4271  \ \ \ \ \ \     \ \ \  &33.1   \ \ \ \ \ \     \ \ \ & 2.36 \ \ \ \ \ \     \ \ \  &$s$   \ \ \ \ \ \     \ \ \  &294 \ \ \ \ \ \     \ \ \  &9.61  \ \ \ \ \ \     \ \ \ & 35.7  \\
$|\Omega_{cc}$$^2P_{\lambda}\frac{3}{2}^-\rangle$  \ \ \ \ \ \     \ \ \  & 4325  \ \ \ \ \ \     \ \ \  &11.4   \ \ \ \ \ \     \ \ \  &174   \ \ \ \ \ \     \ \ \ &$s$  \ \ \ \ \ \     \ \ \  &430  \ \ \ \ \ \     \ \ \ &157  \ \ \ \ \ \     \ \ \  &185  \\
$|\Omega_{cc}$$^4P_{\lambda}\frac{1}{2}^-\rangle$  \ \ \ \ \ \     \ \ \ & 4208  \ \ \ \ \ \     \ \ \ &323 \ \ \ \ \ \ \  \ \ \ &$\cdot\cdot\cdot$   \ \ \ \    \ \ \ &$s$   \ \ \ \ \ \     \ \ \  &3.19  \ \ \ \ \ \     \ \ \  &209  \ \ \ \ \ \     \ \ \  & 323  \\
$|\Omega_{cc}$$^4P_{\lambda}\frac{3}{2}^-\rangle$  \ \ \ \ \ \     \ \ \  & 4252 \ \ \ \ \ \     \ \ \  &3.08  \ \ \ \ \ \     \ \ \ &137   \ \ \ \ \ \  \ \ \ &$s$   \ \ \ \ \ \     \ \ \  &12.9  \ \ \ \ \ \     \ \ \  &202  \ \ \ \ \ \     \ \ \ &140  \\
$|\Omega_{cc}$$^4P_{\lambda}\frac{5}{2}^-\rangle$  \ \ \ \ \ \     \ \ \  & 4303 \ \ \ \ \ \     \ \ \  &41.5 \ \ \ \ \ \     \ \ \ &4.38  \ \ \ \ \ \     \ \ \  &$s$   \ \ \ \ \ \     \ \ \ &12.0  \ \ \ \ \ \     \ \ \  &225   \ \ \ \ \ \     \ \ \ &45.9  \\
\hline\hline
\end{tabular}
\end{table*}
%\end{center}
%\end{widetext}

\begin{figure}[t]
\centering \epsfxsize=8.0 cm \epsfbox{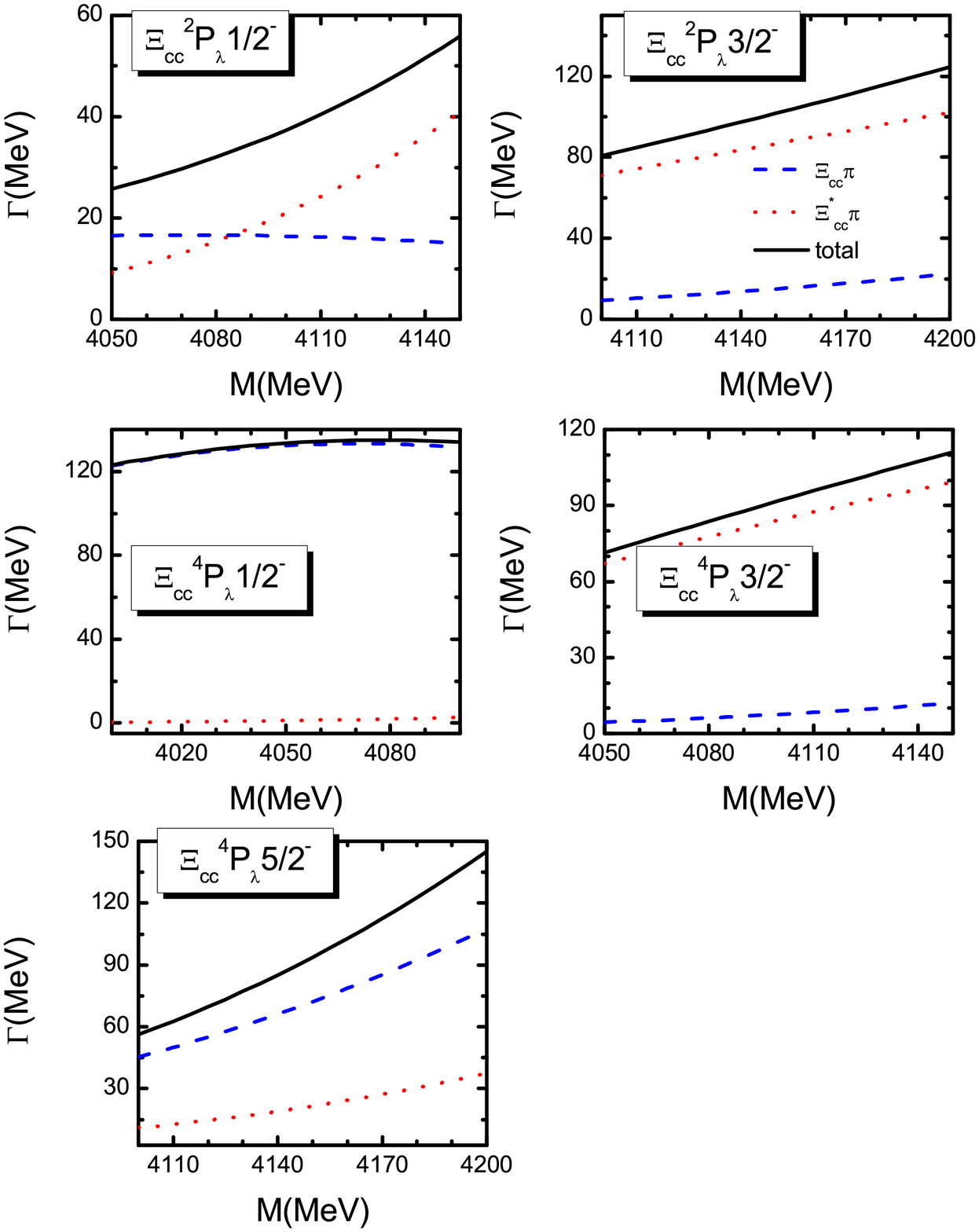}
\caption{(Color online) The strong decay partial widths of the
$1P_{\lambda}$-wave $\Xi_{cc}$ states as a function of the mass.
}\label{fig-Xps}
\end{figure}

\begin{figure}[t]
\centering \epsfxsize=7.75 cm \epsfbox{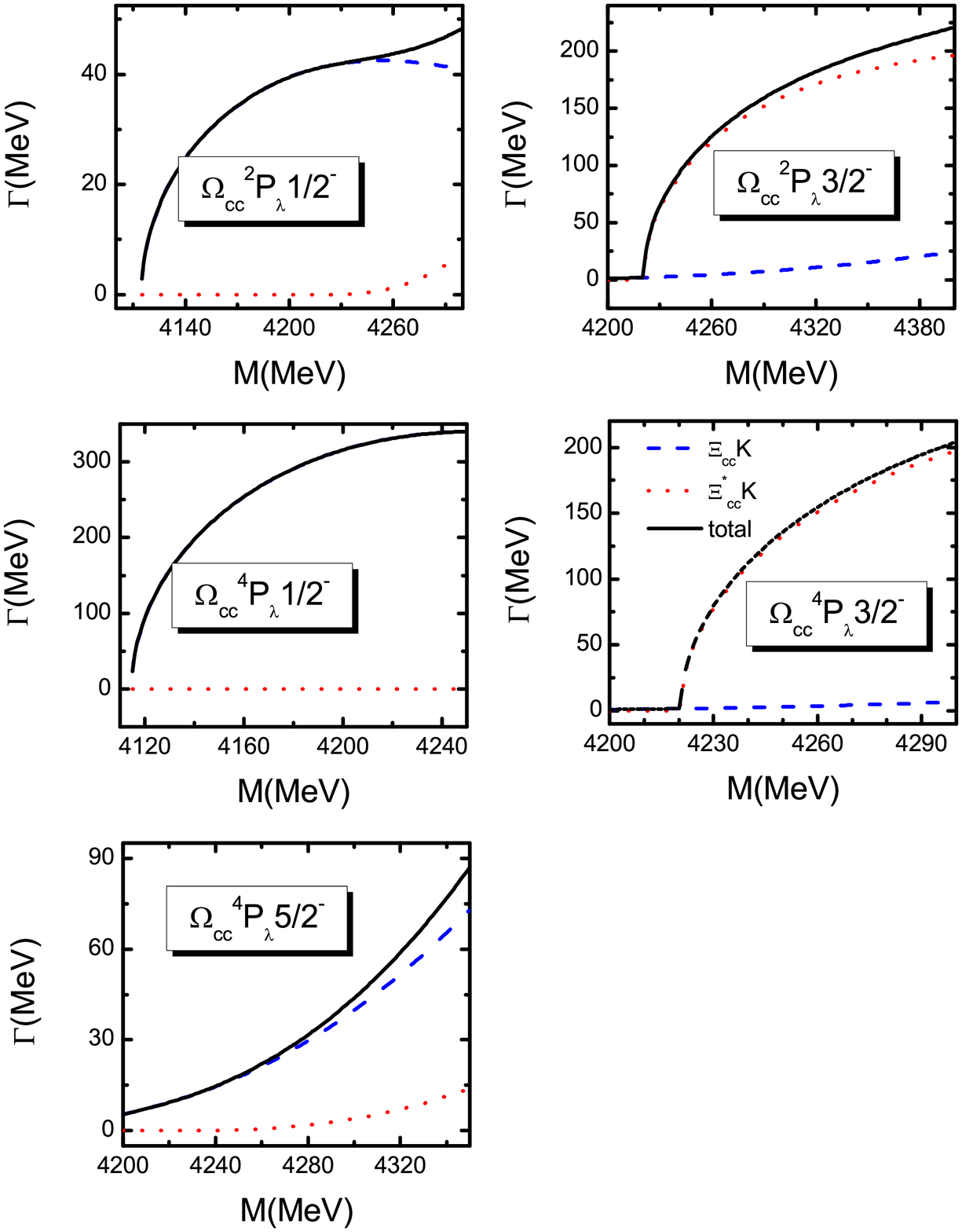}
\caption{(Color online) The strong decay partial widths of the
$1P_{\lambda}$-wave $\Omega_{cc}$ states as a function of the mass.
}\label{fig-Ops}
\end{figure}

%\begin{widetext}
%\begin{center}
\begin{table*}[]
\caption{The variation of the strong decay width of the
$1P_{\lambda}$ states with the harmonic oscillator parameter
$\alpha_{\rho}$. $\Gamma_{\text{total}}$ is the total decay width.
The unit of the mass, width, and $\alpha_{\rho}$ is MeV. }
\label{av}
\begin{tabular}{cccccccc}\hline\hline
 State \ \ \ \ \ \ \ \ \ & Mass\ \ \ \ \ \ \ \ \ & Width
 \ \ \ \ \ \ \ \ \ &$\alpha_{\rho}=620$\ \ \ \ \ \ \ \ \ &$\alpha_{\rho}=580$\ \ \ \ \ \ \ \ \ &$\alpha_{\rho}=540$ \ \ \ \ \ \ \ \ \ &$\alpha_{\rho}=500$\ \ \ \ \ \ \ \ \  &$\alpha_{\rho}=460$  \\
$|\Xi_{cc}\ ^2P_{\lambda}\frac{1}{2}^-\rangle$ \ \ \ \ \ \ \ \ \ &4136\ \ \ \ \ \ \ \ \ & $\Gamma[\Xi_{cc}\pi]$ \ \ \ \ \   \ \ \ & 9.80\ \ \ \ \  \ \ \ &5.39 \ \ \ \ \ \ \ \ &2.34\ \ \ \ \ \ \ \ & 0.58\ \ \ \ \   \ \ \ &0.02 \\
 \ \ \ \ \ \ \ \ \ &\ \ \ \ \ \ \ \ \ & $\Gamma[\Xi^*_{cc}\pi]$ \ \ \ \ \   \ \ \ & 37.2\ \ \ \ \  \ \ \ &40.2 \ \ \ \ \ \ \ \ &43.3\ \ \ \ \ \ \ \ & 46.4\ \ \ \ \   \ \ \ &49.2 \\
\ \ \ \ \ \ \ \ \ &\ \ \ \ \ \ \ \ \ & $\Gamma_{\text{total}}$ \ \ \ \ \   \ \ \ & 47.0\ \ \ \ \  \ \ \ &45.6 \ \ \ \ \ \ \ \ &45.7\ \ \ \ \ \ \ \ & 47.0\ \ \ \ \   \ \ \ &49.2 \\
$|\Xi_{cc}\ ^2P_{\lambda}\frac{3}{2}^-\rangle$ \ \ \ \ \ \ \ \ \ &4196\ \ \ \ \ \ \ \ \ & $\Gamma[\Xi_{cc}\pi]$ \ \ \ \ \   \ \ \ & 22.5\ \ \ \ \  \ \ \ &23.0 \ \ \ \ \ \ \ \ &23.1\ \ \ \ \ \ \ \ & 22.8\ \ \ \ \   \ \ \ &21.6 \\
 \ \ \ \ \ \ \ \ \ &\ \ \ \ \ \ \ \ \ & $\Gamma[\Xi^*_{cc}\pi]$ \ \ \ \ \   \ \ \ & 81.8\ \ \ \ \  \ \ \ &66.8 \ \ \ \ \ \ \ \ &55.8\ \ \ \ \ \ \ \ & 48.6\ \ \ \ \   \ \ \ &44.8 \\
\ \ \ \ \ \ \ \ \ &\ \ \ \ \ \ \ \ \ & $\Gamma_{\text{total}}$ \ \ \ \ \   \ \ \ & 104\ \ \ \ \  \ \ \ &89.8 \ \ \ \ \ \ \ \ &78.9\ \ \ \ \ \ \ \ & 71.3\ \ \ \ \   \ \ \ &66.3 \\
$|\Xi_{cc}\ ^4P_{\lambda}\frac{1}{2}^-\rangle$ \ \ \ \ \ \ \ \ \ &4053\ \ \ \ \ \ \ \ \ & $\Gamma[\Xi_{cc}\pi]$ \ \ \ \ \   \ \ \ & 94.6\ \ \ \ \  \ \ \ &62.7 \ \ \ \ \ \ \ \ &37.2\ \ \ \ \ \ \ \ & 18.4\ \ \ \ \   \ \ \ &6.26 \\
 \ \ \ \ \ \ \ \ \ &\ \ \ \ \ \ \ \ \ & $\Gamma[\Xi^*_{cc}\pi]$ \ \ \ \ \   \ \ \ & 1.34\ \ \ \ \  \ \ \ &1.48 \ \ \ \ \ \ \ \ &1.64\ \ \ \ \ \ \ \ & 1.82\ \ \ \ \   \ \ \ &2.01 \\
\ \ \ \ \ \ \ \ \ &\ \ \ \ \ \ \ \ \ & $\Gamma_{\text{total}}$ \ \ \ \ \   \ \ \ & 95.9\ \ \ \ \  \ \ \ &64.2 \ \ \ \ \ \ \ \ &38.8\ \ \ \ \ \ \ \ & 20.2\ \ \ \ \   \ \ \ &8.27 \\
$|\Xi_{cc}\ ^4P_{\lambda}\frac{3}{2}^-\rangle$ \ \ \ \ \ \ \ \ \ &4101\ \ \ \ \ \ \ \ \ & $\Gamma[\Xi_{cc}\pi]$ \ \ \ \ \   \ \ \ & 8.11\ \ \ \ \  \ \ \ &8.58 \ \ \ \ \ \ \ \ &9.00\ \ \ \ \ \ \ \ & 9.32\ \ \ \ \   \ \ \ &9.49 \\
 \ \ \ \ \ \ \ \ \ &\ \ \ \ \ \ \ \ \ & $\Gamma[\Xi^*_{cc}\pi]$ \ \ \ \ \   \ \ \ & 67.0\ \ \ \ \  \ \ \ &51.7 \ \ \ \ \ \ \ \ &39.0\ \ \ \ \ \ \ \ & 29.0\ \ \ \ \   \ \ \ &22.5 \\
\ \ \ \ \ \ \ \ \ &\ \ \ \ \ \ \ \ \ & $\Gamma_{\text{total}}$ \ \ \ \ \   \ \ \ & 75.1\ \ \ \ \  \ \ \ &60.3 \ \ \ \ \ \ \ \ &48.0\ \ \ \ \ \ \ \ & 38.2\ \ \ \ \   \ \ \ &31.0 \\
$|\Xi_{cc}\ ^4P_{\lambda}\frac{5}{2}^-\rangle$ \ \ \ \ \ \ \ \ \ &4155\ \ \ \ \ \ \ \ \ & $\Gamma[\Xi_{cc}\pi]$ \ \ \ \ \   \ \ \ & 78.9\ \ \ \ \  \ \ \ &81.9 \ \ \ \ \ \ \ \ &83.9\ \ \ \ \ \ \ \ & 84.5\ \ \ \ \   \ \ \ &82.7 \\
 \ \ \ \ \ \ \ \ \ &\ \ \ \ \ \ \ \ \ & $\Gamma[\Xi^*_{cc}\pi]$ \ \ \ \ \   \ \ \ & 24.6\ \ \ \ \  \ \ \ &26.4 \ \ \ \ \ \ \ \ &28.3\ \ \ \ \ \ \ \ & 30.0\ \ \ \ \   \ \ \ &31.5 \\
\ \ \ \ \ \ \ \ \ &\ \ \ \ \ \ \ \ \ & $\Gamma_{\text{total}}$ \ \ \ \ \   \ \ \ & 103\ \ \ \ \  \ \ \ &108 \ \ \ \ \ \ \ \ &112\ \ \ \ \ \ \ \ & 114\ \ \ \ \   \ \ \ &114 \\
\hline\hline
$|\Omega_{cc}\ ^2P_{\lambda}\frac{1}{2}^-\rangle$ \ \ \ \ \ \ \ \ \ &4271\ \ \ \ \ \ \ \ \ & $\Gamma[\Xi_{cc}K]$ \ \ \ \ \   \ \ \ & 32.4\ \ \ \ \  \ \ \ &23.8 \ \ \ \ \ \ \ \ &16.4\ \ \ \ \ \ \ \ & 10.4\ \ \ \ \   \ \ \ &5.71 \\
 \ \ \ \ \ \ \ \ \ &\ \ \ \ \ \ \ \ \ & $\Gamma[\Xi^*_{cc}K]$ \ \ \ \ \   \ \ \ & 2.64\ \ \ \ \  \ \ \ &2.98 \ \ \ \ \ \ \ \ &3.37\ \ \ \ \ \ \ \ & 3.84\ \ \ \ \   \ \ \ &4.41 \\
\ \ \ \ \ \ \ \ \ &\ \ \ \ \ \ \ \ \ & $\Gamma_{\text{total}}$ \ \ \ \ \   \ \ \ & 35.1\ \ \ \ \  \ \ \ &26.8 \ \ \ \ \ \ \ \ &19.8\ \ \ \ \ \ \ \ & 14.2\ \ \ \ \   \ \ \ &10.1 \\
$|\Omega_{cc}\ ^2P_{\lambda}\frac{3}{2}^-\rangle$ \ \ \ \ \ \ \ \ \ &4325\ \ \ \ \ \ \ \ \ & $\Gamma[\Xi_{cc}K]$ \ \ \ \ \   \ \ \ & 12.2\ \ \ \ \  \ \ \ &13.0 \ \ \ \ \ \ \ \ &13.8\ \ \ \ \ \ \ \ & 14.4\ \ \ \ \   \ \ \ &14.8 \\
 \ \ \ \ \ \ \ \ \ &\ \ \ \ \ \ \ \ \ & $\Gamma[\Xi^*_{cc}K]$ \ \ \ \ \   \ \ \ & 143\ \ \ \ \  \ \ \ &115 \ \ \ \ \ \ \ \ &90.7\ \ \ \ \ \ \ \ & 69.4\ \ \ \ \   \ \ \ &51.6 \\
\ \ \ \ \ \ \ \ \ &\ \ \ \ \ \ \ \ \ & $\Gamma_{\text{total}}$ \ \ \ \ \   \ \ \ & 155\ \ \ \ \  \ \ \ &128 \ \ \ \ \ \ \ \ &104\ \ \ \ \ \ \ \ & 83.8\ \ \ \ \   \ \ \ &66.4 \\
$|\Omega_{cc}\ ^4P_{\lambda}\frac{1}{2}^-\rangle$ \ \ \ \ \ \ \ \ \ &4208\ \ \ \ \ \ \ \ \ & $\Gamma[\Xi_{cc}K]$ \ \ \ \ \   \ \ \ & 264\ \ \ \ \  \ \ \ &211 \ \ \ \ \ \ \ \ &163\ \ \ \ \ \ \ \ & 120\ \ \ \ \   \ \ \ &83.3 \\
 \ \ \ \ \ \ \ \ \ &\ \ \ \ \ \ \ \ \ & $\Gamma[\Xi^*_{cc}K]$ \ \ \ \ \   \ \ \ &$\cdot\cdot\cdot$ \ \ \ \ \  \ \ \ &$\cdot\cdot\cdot$ \ \ \ \ \ \ \ \ &$\cdot\cdot\cdot$\ \ \ \ \ \ \ \ &$\cdot\cdot\cdot$ \ \ \ \ \   \ \ \ &$\cdot\cdot\cdot$ \\
\ \ \ \ \ \ \ \ \ &\ \ \ \ \ \ \ \ \ & $\Gamma_{\text{total}}$ \ \ \ \ \   \ \ \ & 264\ \ \ \ \  \ \ \ &211 \ \ \ \ \ \ \ \ &163\ \ \ \ \ \ \ \ & 120\ \ \ \ \   \ \ \ &83.3 \\
$|\Omega_{cc}\ ^4P_{\lambda}\frac{3}{2}^-\rangle$ \ \ \ \ \ \ \ \ \ &4252\ \ \ \ \ \ \ \ \ & $\Gamma[\Xi_{cc}K]$ \ \ \ \ \   \ \ \ & 3.36\ \ \ \ \  \ \ \ &3.68 \ \ \ \ \ \ \ \ &4.02\ \ \ \ \ \ \ \ &4.38\ \ \ \ \   \ \ \ &4.75 \\
 \ \ \ \ \ \ \ \ \ &\ \ \ \ \ \ \ \ \ & $\Gamma[\Xi^*_{cc}K]$ \ \ \ \ \   \ \ \ &118 \ \ \ \ \  \ \ \ &100 \ \ \ \ \ \ \ \ &84.3\ \ \ \ \ \ \ \ &69.3 \ \ \ \ \   \ \ \ &55.6 \\
\ \ \ \ \ \ \ \ \ &\ \ \ \ \ \ \ \ \ & $\Gamma_{\text{total}}$ \ \ \ \ \   \ \ \ & 121\ \ \ \ \  \ \ \ &104 \ \ \ \ \ \ \ \ &88.3\ \ \ \ \ \ \ \ & 73.7\ \ \ \ \   \ \ \ &60.4 \\
$|\Omega_{cc}\ ^4P_{\lambda}\frac{5}{2}^-\rangle$ \ \ \ \ \ \ \ \ \ &4303\ \ \ \ \ \ \ \ \ & $\Gamma[\Xi_{cc}K]$ \ \ \ \ \   \ \ \ & 44.7\ \ \ \ \  \ \ \ &48.0 \ \ \ \ \ \ \ \ &51.3\ \ \ \ \ \ \ \ & 54.3\ \ \ \ \   \ \ \ &56.8 \\
 \ \ \ \ \ \ \ \ \ &\ \ \ \ \ \ \ \ \ & $\Gamma[\Xi^*_{cc}K]$ \ \ \ \ \   \ \ \ &4.86 \ \ \ \ \  \ \ \ &5.41 \ \ \ \ \ \ \ \ &6.05\ \ \ \ \ \ \ \ &6.79 \ \ \ \ \   \ \ \ &7.63 \\
\ \ \ \ \ \ \ \ \ &\ \ \ \ \ \ \ \ \ & $\Gamma_{\text{total}}$ \ \ \ \ \   \ \ \ & 49.6\ \ \ \ \  \ \ \ &53.4 \ \ \ \ \ \ \ \ &57.3\ \ \ \ \ \ \ \ & 61.1\ \ \ \ \   \ \ \ &64.4 \\
\hline\hline
\end{tabular}
\end{table*}
%\end{center}
%\end{widetext}

\section{Calculations and Results }\label{results}

Since the predicted mass of the lowest doubly charmed baryon
$\Xi_{cc}(3621)^{++}$ in relativistic quark model
~\cite{Ebert:2002ig} agrees with the recent experimental measurement
by the LHCb collaboration, we adopt the masses of the doubly
charmed states from Ref.~\cite{Ebert:2002ig} (see Table~\ref{mass})
in our calculation.

\subsection{ The ground doubly charmed states with $J^P=3/2^+$ }

The ground-state $|\Xi_{cc}$$^4S\frac{3}{2}^+\rangle$ has a mass
near $M\sim3727$ MeV ~\cite{Ebert:2002ig}, which is below the
$\Xi_{cc}\pi$ threshold. Thus, the two-body strong decays are
forbidden. This state should mainly decay into $\Xi_{cc}\gamma$. We
plot the radiative decay widths of
$|\Xi_{ccu}$$^4S\frac{3}{2}^+\rangle$ and
$|\Xi_{ccd}$$^4S\frac{3}{2}^+\rangle$ (denoted with $\Xi_{cc}^{*++}$
and $\Xi_{cc}^{*+}$, respectively) as a function of their masses in
Figs.~\ref{fig-Xpr1} and~\ref{fig-Xpr2}. The radiative decay widths
are sensitive to the parent baryon masses. With $M$=3727 MeV, the
radiative partial decay widths of $\Xi_{cc}^{*++}$ and
$\Xi_{cc}^{*+}$ are
\begin{eqnarray}
\Gamma[\Xi_{cc}^{*++}\to \Xi_{cc}^{++}\gamma]&\simeq & 16.7 ~\text{keV},\\
\Gamma[\Xi_{cc}^{*+} \to\Xi_{cc}^+\gamma]&\simeq& 14.6 ~\text{keV},
\end{eqnarray}
respectively, which are comparable with the predictions in
Refs.~\cite{Branz:2010pq} and ~\cite{Li:2017} (see Tab.~\ref{Srad}). The fairly large radiative partial decay
widths indicate the missing $J^P=3/2^+$ ground states
$\Xi_{cc}^{*++}$ ($\Xi_{cc}^{*+}$) might be observed in the
$\Xi_{cc}^{++}\gamma$ ($\Xi_{cc}^{+}\gamma$) channel.

The predicted mass of ground-state
$|\Omega_{cc}$$^4S\frac{3}{2}^+\rangle$ (denoted with
$\Omega_{cc}^*$) is around 3.87 GeV~\cite{Ebert:2002ig}, which is
obviously below the $\Xi_{cc}K$ threshold. This state mainly decays through
the EM transition. From Fig.~\ref{fig-Opr}, the $\Omega_{cc}^*$ has
a quite narrow radiative decay width of $\Gamma\simeq (0-2)$ keV if
the mass of $\Omega_{cc}^*$ is less than 3.84 GeV. With $M=3872$
MeV, we obtain
\begin{equation}
\Gamma(\Omega_{cc}^*\to \Omega_{cc}\gamma)\simeq6.93 ~\mathrm{keV},
\end{equation}
which is compatible with the results in Refs.~\cite{Branz:2010pq} and ~\cite{Li:2017}.

\subsection{The  $P$-wave doubly charmed states}

\subsubsection{$\rho$-mode excitations}

As emphasized in Sec.~\ref{model}, the $\rho$-mode orbitally excited
state has relatively smaller mass than a $\lambda$-mode orbitally
excited state for the doubly heavy baryon system. The masses of the
$1P_{\rho}$ states of $\Xi_{cc}$ and $\Omega_{cc}$ are above the
threshold of $\Xi_{cc}\pi$ and $\Xi_{cc}K$, respectively. However,
their strong decays are forbidden due to the orthogonality of
spatial wave functions if we adopt the simple harmonic oscillator
wave functions for the $1P$ and $1S$ states. In present work we
focus on their radiative decays.

In the quark model, there are two $1P_{\rho}$ states with
$J^P=1/2^-$ and $3/2^-$, respectively. In the doubly charmed baryon
$\Xi_{cc}$ family, the predicted mass of
$|^2P_{\rho}\frac{1}{2}^-\rangle$ is about 3.84
GeV~\cite{Ebert:2002ig}. The mass of
$|^2P_{\rho}\frac{3}{2}^-\rangle$ is about 3.96
GeV~\cite{Ebert:2002ig}, which is $\sim120$ MeV heavier than that of
$|^2P_{\rho}\frac{1}{2}^-\rangle$. Considering the uncertainties of
the predicted mass, we also plot the radiative decay width of
$1P_{\rho}$ as a functions of the parent baryon mass in
Figs.~\ref{fig-Xpr1} and~\ref{fig-Xpr2}. It is interesting to note
that the radiative decay widths of the $\rho$-mode orbitally excited
states are isospin independent because the decay amplitude does not
depend on the light quark. The $|^2P_{\rho}\frac{1}{2}^-\rangle$
should be a narrow state with a width of $\Gamma\leq0.5$ keV. The
state $|^2P_{\rho}\frac{3}{2}^-\rangle$ has a decay width of
$\Gamma\leq2$ keV. Since their strong decays are forbidden, the
total decay widths are almost saturated by the total radiative decay
widths. With the masses of $|^2P_{\rho}\frac{1}{2}^-\rangle$ and
$|^2P_{\rho}\frac{3}{2}^-\rangle$ $M=3838$ MeV and $M=3959$ MeV,
respectively, their total decay widths are
\begin{equation}
\Gamma^{\text{total}}_{|^2P_{\rho}\frac{1}{2}^-\rangle}\simeq0.15
~\text{keV},~~~~~~~~~\Gamma^{\text{total}}_{|^2P_{\rho}\frac{3}{2}^-\rangle}\simeq1.43
~\text{keV}.
\end{equation}

In the $\Omega_{cc}$ family, the masses of
$|^2P_{\rho}\frac{1}{2}^-\rangle$ and
$|^2P_{\rho}\frac{3}{2}^-\rangle$ are about 4.00 and 4.10
GeV~\cite{Ebert:2002ig} respectively. From Fig.~\ref{fig-Opr}, the
$|^2P_{\rho}\frac{1}{2}^-\rangle$ has a total decay width of
$\Gamma\leq0.5$ keV and $|^2P_{\rho}\frac{3}{2}^-\rangle$ has a
total decay width of $\Gamma\leq3.0$ keV, and their main decay
channel is $\Omega_{cc}\gamma$.

\subsubsection{$\lambda$-mode excitations}

In the $\Xi_{cc}$ family, the mass of the first orbital excitation
of the $\lambda$-mode ($1P_{\lambda}$ states) is about 4.10
GeV~\cite{Ebert:2002ig}. The total decay width of
$|\Xi_{cc}~^2P_{\lambda}\frac{1}{2}^-\rangle$ is $\Gamma\simeq 50$
MeV. The main decay channels are $\Xi_{cc}\pi$ and $\Xi^*_{cc}\pi$.
The partial width ratio between $\Xi_{cc}\pi$ and $\Xi^*_{cc}\pi$ is
\begin{equation}
\frac{\Gamma[|\Xi_{cc}~^2P_{\lambda}\frac{1}{2}^-\rangle\to\Xi_{cc}\pi]}{\Gamma[|\Xi_{cc}~^2P_{\lambda}\frac{1}
{2}^-\rangle\to\Xi^*_{cc}\pi]}\simeq0.46.
\end{equation}
On the other hand, the radiative decay rate of
$|\Xi_{cc}^{++}~^2P_{\lambda}\frac{1}{2}^-\rangle$ into
$\Xi^{*++}_{cc}\gamma$ is large, and the predicted branching
fraction is
\begin{equation}
\mathcal{B}\left[|\Xi_{cc}^{++}~^2P_{\lambda}1/2^-\rangle\to
\Xi^{*++}_{cc}\gamma\right]\simeq0.24\%.
\end{equation}
The radiative decay rate of
$|\Xi_{cc}^{+}~^2P_{\lambda}\frac{1}{2}^-\rangle$ into
$\Xi^{+}_{cc}\gamma$ is significant, and the predicted branching
ratio is
\begin{equation}
\mathcal{B}\left[|\Xi_{cc}^{+}~^2P_{\lambda}1/2^-\rangle\to
\Xi^{+}_{cc}\gamma\right]\simeq0.50\%.
\end{equation}

The states $|\Xi_{cc}$$^2P_{\lambda}\frac{3}{2}^-\rangle$ and
$|\Xi_{cc}$$^4P_{\lambda}\frac{3}{2}^-\rangle$ have a moderate width
of $\Gamma\sim100$ MeV, and dominantly decay into $\Xi^*_{cc}\pi$.
The partial width of
$\Gamma[|\Xi_{cc}$$^2P_{\lambda}\frac{3}{2}^-\rangle\to
\Xi_{cc}\pi]$ is sizable. The partial width ratio is
\begin{eqnarray}
\frac{\Gamma[|\Xi_{cc}~^2P_{\lambda}3/2^-\rangle\rightarrow
\Xi_{cc}\pi]} {\Gamma[|\Xi_{cc}~^2P_{\lambda}3/2^-\rangle\rightarrow
\Xi^*_{cc}\pi]}\simeq 0.21,
\end{eqnarray}
which can be used to distinguish
$|\Xi_{cc}$$^2P_{\lambda}\frac{3}{2}^-\rangle$ from
$|\Xi_{cc}$$^4P_{\lambda}\frac{3}{2}^-\rangle$ in future
experiments. The radiative partial widths of the $P$-wave
$\Xi_{cc}^{++}$ states with $J^P=3/2^-$ into $\Xi_{cc}^{++}\gamma$
and $\Xi_{cc}^{*++}\gamma$ are around a few hundred keV. The
radiative decays of $|\Xi_{cc}^+$$^2P_{\lambda}\frac{3}{2}^-\rangle$
and $|\Xi_{cc}^+$$^4P_{\lambda}\frac{3}{2}^-\rangle$ are dominated
by $\Xi_{cc}^+\gamma$ and $\Xi_{cc}^{*+}\gamma$, respectively. Their
partial width can also reach up to several hundred keV as well.
These radiative processes may be measured in future experiments due
to their sizeable branching fractions, $\mathcal{O}(10^{-3})$.

From the Table~\ref{PLV}, the decay widths of
$|\Xi_{cc}$$^4P_{\lambda}\frac{1}{2}^-\rangle$ and
$|\Xi_{cc}$$^4P_{\lambda}\frac{5}{2}^-\rangle$ are about
$\Gamma\sim100$ MeV. The dominant decay mode of
$|\Xi_{cc}$$^4P_{\lambda}\frac{1}{2}^-\rangle$ is $\Xi_{cc}\pi$,
while $|\Xi_{cc}$$^4P_{\lambda}\frac{5}{2}^-\rangle$ mainly decays
into $\Xi_{cc}\pi$ and $\Xi^*_{cc}\pi$ channels with the partial
decay ratio
\begin{eqnarray}
\frac{\Gamma[|\Xi_{cc}~^4P_{\lambda}5/2^-\rangle\rightarrow
\Xi^*_{cc}\pi]}
{\Gamma[|\Xi_{cc}~^4P_{\lambda}5/2^-\rangle\rightarrow
\Xi_{cc}\pi]}\simeq 0.30.
\end{eqnarray}
For the EM transitions, the $\Xi^*_{cc}\gamma$ channel is their
dominant decay mode. The radiative decay partial widths into
$\Xi^{*++,+}_{cc}\gamma$ are a few hundred keV. The branching
fractions for these radiative decay processes are
$\mathcal{O}(10^{-3})$. These sizeable branching fractions indicate
that the radiative decays of
$|\Xi_{cc}$$^4P_{\lambda}\frac{1}{2}^-\rangle$ and
$|\Xi_{cc}$$^4P_{\lambda}\frac{5}{2}^-\rangle$ may be observed in
future experiments.

We analyze the decay properties of the $1P_{\lambda}$ states in the
$\Omega_{cc}$ family and collect their partial strong and radiative
decay widths in Table~\ref{PLV}. The states
$|\Omega_{cc}$$^2P_{\lambda}\frac{1}{2}^-\rangle$ and
$|\Omega_{cc}$$^4P_{\lambda}\frac{5}{2}^-\rangle$ are most likely to
be the narrow states with a width of $\Gamma\sim40$ MeV, and the
$\Xi_{cc}K$ decay channel almost saturates their total decay widths.
The dominant radiative decay modes of these two states are
$\Omega_{cc}\gamma$ and $\Omega^*_{cc}\gamma$, respectively. The
branching ratios are
\begin{eqnarray}
\mathcal{B}[^2P_{\lambda}1/2^-\rightarrow \Omega_{cc}\gamma]\simeq0.81\%,\\
\mathcal{B}[^4P_{\lambda}5/2^-\rightarrow
\Omega^*_{cc}\gamma]\simeq0.48\%,
\end{eqnarray}
which can be tested in future experiments.

The state $|\Omega_{cc}$$^4P_{\lambda}\frac{1}{2}^-\rangle$ has a
broad width of $\Gamma\simeq320$ MeV, and mainly decays into
$\Xi_{cc}K$. The states
$|\Omega_{cc}$$^2P_{\lambda}\frac{3}{2}^-\rangle$ and
$|\Omega_{cc}$$^4P_{\lambda}\frac{3}{2}^-\rangle$ may lie below the
threshold of $\Xi^*_{cc}K$. If so, they mainly decay into
$\Xi_{cc}K$ channel with a fairly narrow width $\Gamma\sim11$ MeV
and $\Gamma\sim3$ MeV, respectively. The branching ratios of the
main radiative decay channels are
\begin{eqnarray}
\mathcal{B}[^2P_{\lambda}3/2^-\rightarrow \Omega_{cc}\gamma]\simeq3.77\%,\\
\mathcal{B}[^4P_{\lambda}3/2^-\rightarrow
\Omega^*_{cc}\gamma]\simeq6.49\%,
\end{eqnarray}
respectively, which are significant and can be searched for in
experiment. However, if their masses are above the threshold of
$\Xi^*_{cc}K$, their dominant decay modes should be $\Xi^*_{cc}K$
and their total decay widths may reach $\Gamma\sim185$ MeV and
$\Gamma\sim140$ MeV.

Considering the uncertainty of the mass predictions of the
$1P_{\lambda}$ states, we plot the radiative and strong decay widths
as a function of the mass in Figs.~\ref{fig-Xpr1}-~\ref{fig-Ops},
respectively. The sensitivities of the decay properties of the
doubly charmed states to their masses can be clearly seen from these
figures.

\subsection{The effect of $\alpha_{\rho}$}

We have discussed the decay properties of the ground state with
$J^P=3/2^+$ and $1P$-wave states with $J^P=1/2^-$, $3/2^-$, $5/2^-$
for the doubly charmed baryons. All of the theoretical predictions
in the present work are obtained with the parameter
$\alpha_{\rho}=660$ MeV. However, the harmonic oscillator parameter
$\alpha_{\rho}$ is not determined absolutely, which bares a large
uncertainty. Fixing the mass values as in Ref.~\cite{Ebert:2002ig},
we further consider the decay properties as a function of the
parameter $\alpha_{\rho}$. The results are listed in Table \ref{av}.

It's important to note that the effects on radiative decay widths
from the parameter $\alpha_{\rho}$ are much smaller than that on
strong decay widths, so we just list the hadronic decay width in
table. From the table, only the decay properties of the doubly
charmed states with $J^P=1/2^-$ are sensitive to the harmonic
oscillator parameter $\alpha_{\rho}$. Fortunately, the decay
properties of the other doubly charmed states with $J^P=3/2^-$ and
$J^P=5/2^-$ states are less sensitivities to this parameter. Thus,
our numerical results and main predictions in present work should
hold in a reasonable range of the parameter $\alpha_{\rho}$.

%\begin{widetext}
%\begin{center}
\begin{table*}
\caption{The partial widths of strong and radiative decays for the $1P$ states.
$\Gamma_{\text{total}}$ stands for the total decay width. } \label{bb}
\begin{tabular}{cccccccc}
\hline\hline
\ \ \ \ \ \  \ \ \ &\ \ \ \ \ \ \ &     \ \ \ &   \ \  & \text{Types of} \ \ \ \ \ \ \ \ \ \ \ & \ \ \ \  \ \ \ &   \ \ \ \    \ \ \ &       \\
State \ \ \ \ \    &  Mass(MeV) \ \ \ \ \ \   & $\Gamma[\Xi_{bb}\pi]$(MeV) \ \ \ \ \ \  & $\Gamma[\Xi^*_{bb}\pi]$(MeV) \ \ \ \ \   & \text{light quark}    \ \   \ \ \ & $\Gamma[\Xi_{bb}\gamma]$(keV) \ \ \ \ \ \   & $\Gamma[\Xi^*_{bb}\gamma]$(keV)  \ \ \ \ \    & $\Gamma_{\text{total}}$(MeV)      \\
$|\Xi_{bb}$$^2P_{\rho}\frac{1}{2}^-\rangle$ \ \ \ \ \ \  \ \ \ & 10368 \ \ \ \ \ \     \ \ \ &$\cdot\cdot\cdot$  \ \ \ \ \ \     \ \ \  &$\cdot\cdot\cdot$ \ \ \ \ \ \     \ \ \  &$u(d)$  \ \ \ \ \ \     \ \ \  &1.15 \ \ \ \ \ \     \ \ \  &0.72   \ \ \ \ \ \     \ \ \ &$1.87\times10^{-3}$  \\
$|\Xi_{bb}$$^2P_{\rho}\frac{3}{2}^-\rangle$  \ \ \ \ \ \     \ \ \ & 10408 \ \ \ \ \ \     \ \ \ &$\cdot\cdot\cdot$  \ \ \ \ \ \     \ \ \ &$\cdot\cdot\cdot$  \ \ \ \ \ \     \ \ \  &$u(d)$  \ \ \ \ \ \     \ \ \ &3.30  \ \ \ \ \ \     \ \ \  &2.66   \ \ \ \ \ \     \ \ \  &$5.96\times10^{-3}$  \\
$|\Xi_{bb}$$^2P_{\lambda}\frac{1}{2}^-\rangle$ \ \ \ \ \ \  \ \ \ & 10675 \ \ \ \ \ \     \ \ \ &1.39  \ \ \ \ \ \     \ \ \  &80.0 \ \ \ \ \ \     \ \ \  &$u$  \ \ \ \ \ \     \ \ \  &455 \ \ \ \ \ \     \ \ \  &235   \ \ \ \ \ \     \ \ \ &82.1  \\
                                                 \ \ \ \ \ \     \ \ \  &        \ \ \ \ \ \     \ \ \  &   \ \ \ \ \ \     \ \ \  &   \ \ \ \ \ \     \ \ \  &$d$  \ \ \ \ \ \     \ \ \  &71.1  \ \ \ \ \ \     \ \ \ &59.3  \ \ \ \ \ \     \ \ \  &     \\
$|\Xi_{bb}$$^2P_{\lambda}\frac{3}{2}^-\rangle$  \ \ \ \ \ \     \ \ \ & 10694 \ \ \ \ \ \     \ \ \ &16.3  \ \ \ \ \ \     \ \ \ &55.4  \ \ \ \ \ \     \ \ \  &$u$  \ \ \ \ \ \     \ \ \ &984  \ \ \ \ \ \     \ \ \  &265   \ \ \ \ \ \     \ \ \  &72.9  \\
                                                 \ \ \ \ \ \     \ \ \  &        \ \ \ \ \ \     \ \ \  &   \ \ \ \ \ \     \ \ \  &  \ \ \ \ \ \     \ \ \  &$d$  \ \ \ \ \ \     \ \ \  &182   \ \ \ \ \ \     \ \ \ &67.1  \ \ \ \ \ \     \ \ \  &       \\
$|\Xi_{bb}$$^4P_{\lambda}\frac{1}{2}^-\rangle$  \ \ \ \ \ \     \ \ \ & 10632  \ \ \ \ \ \     \ \ \  &20.5  \ \ \ \ \ \     \ \ \ &6.16  \ \ \ \ \ \     \ \ \  &$u$   \ \ \ \ \ \     \ \ \ &555  \ \ \ \ \ \     \ \ \  &$1.33\times10^{3}$   \ \ \ \ \ \     \ \ \ &28.5  \\
                                               \ \ \ \ \ \     \ \ \  &    \ \ \ \ \ \     \ \ \  &   \ \ \ \ \ \     \ \ \  &   \ \ \ \ \ \     \ \ \  &$d$  \ \ \ \ \ \     \ \ \  &14.0 \ \ \ \ \ \     \ \ \  &271  \ \ \ \ \ \     \ \ \ &      \\
$|\Xi_{bb}$$^4P_{\lambda}\frac{3}{2}^-\rangle$  \ \ \ \ \ \     \ \ \  & 10647 \ \ \ \ \ \     \ \ \ &8.58    \ \ \ \ \ \     \ \ \  &39.0   \ \ \ \ \ \     \ \ \  &$u$  \ \ \ \ \ \     \ \ \  &172  \ \ \ \ \ \     \ \ \  &773  \ \ \ \ \ \     \ \ \  & 48.5  \\
                                                \ \ \ \ \ \     \ \ \  &    \ \ \ \ \ \     \ \ \  &     \ \ \ \ \ \     \ \ \  &     \ \ \ \ \ \     \ \ \ &$d$  \ \ \ \ \ \     \ \ \  &43.5  \ \ \ \ \ \     \ \ \ &149  \ \ \ \ \ \     \ \ \ &      \\
$|\Xi_{bb}$$^4P_{\lambda}\frac{5}{2}^-\rangle$  \ \ \ \ \ \     \ \ \  & 10661 \ \ \ \ \ \     \ \ \ &58.9  \ \ \ \ \ \     \ \ \ &36.2   \ \ \ \ \ \     \ \ \ &$u$   \ \ \ \ \ \     \ \ \  &121   \ \ \ \ \ \     \ \ \  &569   \ \ \ \ \ \     \ \ \  &95.6  \\
                                                \ \ \ \ \ \     \ \ \  &  \ \ \ \ \ \     \ \ \  &    \ \ \ \ \ \     \ \ \  &     \ \ \ \ \ \     \ \ \  &$d$   \ \ \ \ \ \     \ \ \  &30.5 \ \ \ \ \ \     \ \ \ &104    \ \ \ \ \ \     \ \ \  &      \\
\hline\hline
State \ \ \ \ \    &  Mass(MeV) \ \ \ \ \ \ & $\Gamma[\Xi_{bb}K]$(MeV) \ \ \ \ \ \ & $\Gamma[\Xi^*_{bb}K]$(MeV) \ \ \ \ \ \   &   \ \ \ \ \ \   & $\Gamma[\Omega_{bb}\gamma]$(keV) \ \ \ \ \ \   & $\Gamma[\Omega^*_{bb}\gamma]$(keV)  \ \ \ \ \     & $\Gamma_{\text{total}}$(MeV)      \\
$|\Omega_{bb}$$^2P_{\rho}\frac{1}{2}^-\rangle$  \ \ \ \ \ \     \ \ \  & 10532  \ \ \ \ \ \     \ \ \  &$\cdot\cdot\cdot$   \ \ \ \ \ \     \ \ \ &$\cdot\cdot\cdot$  \ \ \ \ \ \     \ \ \  &$s$   \ \ \ \ \ \     \ \ \  &1.41 \ \ \ \ \ \     \ \ \  &1.11  \ \ \ \ \ \     \ \ \ &$2.52\times10^{-3}$   \\
$|\Omega_{bb}$$^2P_{\rho}\frac{3}{2}^-\rangle$  \ \ \ \ \ \     \ \ \  & 10566  \ \ \ \ \ \     \ \ \  &$\cdot\cdot\cdot$   \ \ \ \ \ \     \ \ \  &$\cdot\cdot\cdot$   \ \ \ \ \ \     \ \ \ &$s$  \ \ \ \ \ \     \ \ \  &3.38  \ \ \ \ \ \     \ \ \ &3.16  \ \ \ \ \ \     \ \ \  &$6.54\times10^{-3}$  \\
$|\Omega_{bb}$$^2P_{\lambda}\frac{1}{2}^-\rangle$  \ \ \ \ \ \     \ \ \  & 10804  \ \ \ \ \ \     \ \ \  &16.2   \ \ \ \ \ \     \ \ \ & 14.1 \ \ \ \ \ \     \ \ \  &$s$   \ \ \ \ \ \     \ \ \  &76.9 \ \ \ \ \ \     \ \ \  &26.2  \ \ \ \ \ \     \ \ \ & 30.4  \\
$|\Omega_{bb}$$^2P_{\lambda}\frac{3}{2}^-\rangle$  \ \ \ \ \ \     \ \ \  & 10821  \ \ \ \ \ \     \ \ \  &6.40   \ \ \ \ \ \     \ \ \  &82.7   \ \ \ \ \ \     \ \ \ &$s$  \ \ \ \ \ \     \ \ \  &151  \ \ \ \ \ \     \ \ \ &30.0  \ \ \ \ \ \     \ \ \  &89.3  \\
$|\Omega_{bb}$$^4P_{\lambda}\frac{1}{2}^-\rangle$  \ \ \ \ \ \     \ \ \ & 10771  \ \ \ \ \ \     \ \ \ &149 \ \ \ \ \ \ \  \ \ \ &0.40 \ \ \ \ \ \     \ \ \  &$s$   \ \ \ \ \ \     \ \ \  &6.38  \ \ \ \ \ \     \ \ \  &188  \ \ \ \ \ \     \ \ \  & 149.6  \\
$|\Omega_{bb}$$^4P_{\lambda}\frac{3}{2}^-\rangle$  \ \ \ \ \ \     \ \ \  & 10785 \ \ \ \ \ \     \ \ \  &2.28  \ \ \ \ \ \     \ \ \ &96.9  \ \ \ \ \ \     \ \ \  &$s$   \ \ \ \ \ \     \ \ \  &20.0  \ \ \ \ \ \     \ \ \  &117  \ \ \ \ \ \     \ \ \ &99.3  \\
$|\Omega_{bb}$$^4P_{\lambda}\frac{5}{2}^-\rangle$  \ \ \ \ \ \     \ \ \  & 10798 \ \ \ \ \ \     \ \ \  &19.0 \ \ \ \ \ \     \ \ \ &6.00 \ \ \ \ \ \     \ \ \  &$s$   \ \ \ \ \ \     \ \ \ &14.2  \ \ \ \ \ \     \ \ \  &90.9   \ \ \ \ \ \     \ \ \ &25.1  \\
\hline\hline
\end{tabular}
\end{table*}
%\end{center}
%\end{widetext}

\subsection{The doubly bottom states}
As a byproduct, we also investigate the strong and radiative decay properties of the ground state with $J^P=3/2^+$ and $1P$-wave states with $J^P=1/2^-$, $3/2^-$, $5/2^-$ for the doubly bottom baryons. Now
the harmonic oscillator parameter $\alpha_{\rho}$ of the $\rho$-mode excitation between the two bottom quarks $\alpha_{\rho}=0.70$ GeV, which is slightly larger than that of the $\rho$-mode excitation between the two charm quarks $\alpha_{\rho}=0.66$ GeV.

We collect the theoretical predictions for the doubly bottom baryons in Tables.~\ref{Srad} and~\ref{bb}. The radiative decay width of the ground states with $J^P=3/2^+$ is quite narrow, which is similar to the predictions in Ref.\cite{Branz:2010pq}. On the other hand, the total decay widths of the $1P_{\lambda}$ states are about several tens MeV.

\section{Summary}\label{suma}

In the framework of the nonrelativistic constituent quark model, we
have systematically studied the strong and radiative decay
properties of the low-lying doubly charmed baryons, i.e., the ground
state with $J^P=3/2^+$ and $1P$-wave states with $J^P=1/2^-$,
$3/2^-$, $5/2^-$. Our main predictions are summarized as follows.

For the ground states with $J^P=3/2^+$, their decays are dominated
by the radiative transitions. The radiative partial width of
$\Xi^{*++}_{cc}$ ($\Xi^{*+}_{cc}$) into $\Xi^{++}_{cc}\gamma$
($\Xi^{+}_{cc}\gamma$) is predicted to be several tens keV. The
$\Xi^{*++}_{cc}$ might be reconstructed in the $\Xi^{++}_{cc}\gamma$
channel with $\Xi^{++}_{cc}\to \Lambda^+_cK^-\pi^+\pi^+$ at LHCb.

We want to emphasize that the lowest lying excited doubly charmed
baryons should be dominated by the $\rho$-mode $1P$-wave components
$|\Xi_{cc}^{++}~^2P_{\rho}\frac{1}{2}^-\rangle$ and
$|\Xi_{cc}^{++}~^2P_{\rho}\frac{3}{2}^-\rangle$, which should be
quite narrow states. Their decay
widths are dominated by the one-photon radiative transitions into
the ground state with $J^P=1/2^+$ due to the absence of the strong
decay modes. In the realistic case, there may exist mixing between the $\rho$-mode and the $\lambda$-mode excitations. Even with the mixing, the ratio of the radiative decays should be significant for the lowest lying excited doubly charmed baryons.
The $|\Xi_{cc}^{++}~^2P_{\rho}\frac{1}{2}^-\rangle$ and
$|\Xi_{cc}^{++}~^2P_{\rho}\frac{3}{2}^-\rangle$ may also be
reconstructed in the $\Xi^{++}_{cc}\gamma$ channel with
$\Xi^{++}_{cc}\to \Lambda^+_cK^-\pi^+\pi^+$ at LHCb.

For the $\lambda$-mode $1P_{\lambda}$ states in $\Xi_{cc}$ family,
their total strong decay widths are about $\Gamma\sim100$ MeV. The
$|\Xi_{cc}^{++}~^4P_{\lambda}\frac{1}{2}^-\rangle$ and
$|\Xi_{cc}^{++}~^4P_{\lambda}\frac{5}{2}^-\rangle$ mainly decay into
the $\Xi^{++}_{cc}\pi^0$ channel, which may be reconstructed with
$\Xi^{++}_{cc}\to \Lambda^+_cK^-\pi^+\pi^+$ at LHCb. The
$|\Xi_{cc}^{++}~^2P_{\lambda}\frac{3}{2}^-\rangle$ and
$|\Xi_{cc}^{++}~^4P_{\lambda}\frac{3}{2}^-\rangle$ mainly decay into
the $\Xi^{*++}_{cc}\pi^0$ channel, which may be searched for by
reconstructing $\Xi^{*++}_{cc}$ in the decay chain of
$\Xi^{*++}_{cc}\to \Xi^{++}_{cc}\gamma \to \Lambda^+_cK^-\pi^+\pi^+
\gamma$ at LHCb.

We have also investigated the strong and radiative decay properties
of the low-lying $S$- and $P$-wave $\Omega_{cc}$, $\Xi_{bb}$, and $\Omega_{bb}$ states in this work.
Hopefully our predictions should be useful to looking for the states
in their corresponding family.

\section*{Acknowledgements }

This work is supported by the National Natural Science Foundation of
China under Grants 11375061, 11575008, 11621131001 and 973 program.

\end{document}